\documentclass{aa}  

\usepackage{graphicx}
\usepackage{txfonts}

\usepackage{xcolor}

\begin{document} 

\title{Modelling two Energetic Storm Particle Events Observed by Solar Orbiter Using the Combined EUHFORIA and iPATH Models}

   \author{Zheyi Ding          \inst{1},
           Gang Li\inst{2}, 
           Glenn Mason \inst{3},
           Stefaan Poedts \inst{1,4},
Athanasios Kouloumvakos \inst{3},
George Ho \inst{3},
Nicolas Wijsen \inst{5,6},
Robert F. Wimmer-Schweingruber\inst{7}, 
Javier Rodríguez-Pacheco
\inst{8}
          }

   \institute{ 
   \inst{1} Centre for mathematical Plasma Astrophysics, KU Leuven, 3001 Leuven, Belgium\\
              \inst{2} Department of Space Science and CSPAR, University of Alabama in Huntsville, Huntsville, AL 35899, USA;
             \email{gangli.uah@gmail.com}\\
              \inst{3} Johns Hopkins Applied Physics Lab, Laurel, MD 20723, USA \\             
             \inst{4} Institute of Physics, University of Maria Curie-Sk{\l}odowska, Pl.\ M.\ Curie-Sk{\l}odowska 5, 20-031 Lublin, Poland \\
            \inst{5} Heliophysics Science Division, NASA Goddard Space Flight Center, Greenbelt, MD 20771, USA\\
            \inst{6} Department of Astronomy, University of Maryland College Park, MD 20742, USA\\
             \inst{7} Institute of Experimental and Applied Physics, University of Kiel, Leibnizstrasse 11, D-24118 Kiel, Germany  \\
             \inst{8} Universidad de Alcalá, Alcalá de Henares 28805, Spain \\   
             }

   \date{Received July 20, 2023; accepted November 10, 2023}

\abstract
{By coupling the EUropean Heliospheric FORcasting Information Asset (EUHFORIA) and the improved Particle Acceleration and Transport in the Heliosphere (iPATH) model, two energetic storm particle (ESP) events, originating from the same active region (AR 13088) and  observed by Solar Orbiter (SolO) on  August 31 2022 and  September 05 2022, are modelled.}
{To better understand particle acceleration and transport process in the inner heliosphere by combining numerical simulations and SolO observations.} 
{We simulate the two coronal mass ejections (CMEs) in a data-driven real-time solar wind background with the EUHFORIA code. MHD parameters concerning the shock and downstream medium are computed from EUHFORIA as inputs for the iPATH model. In the iPATH model, a shell structure is maintained to model the turbulence-enhanced shock sheath. At the shock front, assuming diffuse shock acceleration, the particle distribution is obtained by taking the steady state solution with the instantaneous shock parameters. Upstream of the shock, particles escape, and their transport in the solar wind is described by a focused transport equation using the backward stochastic differential equation method.
}
{While both events originated from the same active region, they exhibited notable differences, including: 1) the August ESP event lasted for 7 hours, while the September event persisted for 16 hours;  2) The time intensity profiles for the September event showed a clear cross-over upstream of the shock  where the intensity of higher energy protons exceeds those of lower energy protons, leading to positive (``reverse'') spectral indices prior to the shock passage. For both events,  our simulations replicate the observed duration of the shock sheath, depending on the deceleration history of the CME.  Imposing different choices of escaping length scale, which is related to the decay of upstream turbulence, the modelled time intensity profiles prior to the shock arrival also agree with observations. In particular, the cross-over of this time profile in the September event is well reproduced. We show that a ``reverse'' upstream spectrum is the result of the interplay between two length scales. One characterizes the decay of upstream shock accelerated particles, which are controlled by the energy-dependent diffusion coefficient, and the other characterizes the decay of  upstream turbulence power, which is related to the process of how streaming protons upstream of the shock excite Alfv\'{e}n waves. }
{The behavior of SEP events depends on many variables. Even similar eruptions from the same AR may lead to SEP events that have very different characteristics. Simulations taking into account real-time background solar wind, the dynamics of the CME propagation, and upstream turbulence at the shock front are necessary to thoroughly understand the ESP phase of large SEP events.}
   \titlerunning {Modelling two ESP events observed by Solar Orbiter}
   \authorrunning{Ding et al.}
   \maketitle

%
\section{introduction}
In large solar energetic particle (SEP) events, protons and ions can be accelerated to very high energies ($>100$ MeV/n), causing a major concern in space weather \citep{Desai2016}.  These SEP events are typically associated with shock waves driven by coronal mass ejections (CMEs), and the acceleration mechanism is thought to be the diffusive shock acceleration (DSA) \citep{Axford1977, Drury+1983}.  DSA predicts a power law spectrum for energetic particles when a steady-state solution is considered. Observations of many SEP events indeed show that power laws can, in general, provide a good description of SEP spectra.  However, a careful examination of individual events shows that there is a large event-to-event variability: events with similar eruption characteristics can vary significantly in maximum particle energy, particle composition, as well as the spectral shape. Even for the same event, different observers connecting to different parts of the shock may see different SEP characteristics  \citep[e.g.,][]{Dresing2014,Mason+2012,Desai2016ions,Kouloumvakos2019ApJ...876...80K}. 
Furthermore, the recent Solar Orbiter (SolO) and Parker Solar Probe (PSP) missions have gathered unprecedented data of energetic particles near the Sun, including some unexpected measurements of SEP events. These measurements offer new clues to better understanding the outstanding problem of charged particle energization and propagation \citep[see the review by][]{Malandraki2023PhPl...30e0501M}. 
These observations suggest that there are still many factors that can affect the particle acceleration process and that require more detailed examination. In-situ observations of SEP events involve the interplay of acceleration and transport: energetic particles are continuously accelerated at the shock front. After being accelerated, they escape upstream and propagate along and cross the interplanetary magnetic field (IMF) lines and reach an observer. The observed ion time profiles and ion spectra therefore are controlled both by the acceleration and the transport process. At the shock passage, there is a phase of the SEP event where local shock parameters and local energetic particles (often signaled by a peak around the shock passage) are observed simultaneously.  This phase is often referred to as an energetic storm particle (ESP) event \citep{Bryant1962}.  It was argued that the study of these events can single out the acceleration process \citep[e.g.,][]{Duenas2022,Lario2018JPhCS1100a2014L,Lario2023ApJ...950...89L,Ding2023JGRA..12831502D}. However, due to enhanced turbulence near the shock complex, downstream of the shock, energetic particles accelerated earlier are trapped within the shock complex. This fact makes the interpretation of downstream particle behavior very difficult.

Our understanding of the acceleration process in ESP events is further complicated by the transport process upstream of the shock. The presence of turbulence in the upstream region is essential for the scattering and escape of energetic particles. It is commonly assumed that the turbulence near the shock takes the form of Alfv\'{e}n waves and upstream of the shock these waves are driven by protons streaming away from the shock front.  This leads to a coupling between the wave intensity and the anisotropy of  particle distribution function \citep{Bell1978MNRAS.182..147B}. 
An earlier attempt to examine the effect of these waves on the particle acceleration process, in the context of interplanetary shocks was taken by \citet{Lee+1983} who solved the coupled particle transport and upstream Alfv\'{e}n wave intensity equations and obtained analytical solutions under the steady-state assumption. Later \cite{Gordon+1999} utilized the steady-state solution for both the  energetic particles  and upstream wave spectra to examine particle acceleration at Earth's bow shock. For traveling shocks such as those driven by CMEs, the particle acceleration is intrinsically a time-dependent problem. \cite{Ng2003ApJ...591..461N} adopted the same set of equations as \cite{Lee+1983} but solved the time-dependent wave transport equation, enabling the determination of wave action and energetic particle spectra in a time-dependent manner. 

Modelling a time-dependent particle acceleration process is itself time-consuming.  One approach,  which was proved successful in modelling large SEP events \citep{Verkhoglyadova+2009,Verkhoglyadova+2010}  
was developed in the Particle Acceleration and Transport in the Heliosphere (PATH) code \citep{Zank+etal+2000, Rice+2003, Li+2003,Li+2005ions}. These authors tracked the propagation of the CME-driven shock numerically and evaluated the instantaneous dynamic time scale of the CME-driven shock.  At
any given time, the solution of the wave intensity, although intrinsically a time-dependent problem, is approximated by a steady-state solution at the shock front, with the maximum particle energy constrained by the instantaneous shock dynamic time scale, as given by \cite{Gordon+1999}. 
Such an approach is further elaborated in the improved PATH (iPATH) model \citep{Hu+etal+2017, Hu+etal+2018} and in the investigation of individual events \citep{Li2021, Ding+2020, ding2022A&A...668A..71D}. In this work, we follow the same approach and combine the iPATH code with the  EUropean Heliospheric FORcasting Information Asset (EUHFORIA; \cite{pomoell2018euhforia}) code.  In contrast to previous work, we do not attempt detailed event fitting. Instead, we pay special attention to various length scales upstream of the shock, and how their relative size can affect the observed features of SEP events. 
Specifically, by comparing two recent ESP events observed by SolO, we aim to understand the interplay between upstream turbulence and the decay of particle intensity upstream of the shock. We analyze the characteristics of upstream magnetic fluctuation and elucidate the underlying mechanisms of particle escape from the shock. These analyses were achieved by utilizing the combined EUHFORIA and iPATH models to simulate the observed time-intensity profiles and spectra.  Our analyses form a basis for understanding different types of ESP phases of SEP events. 

Our paper is organized as follows. In Section~\ref{sec:model}, we briefly explain the coupling between the EUHFORIA and the iPATH model, explaining how shock parameters and shell structures are extracted from EUHFORA and passed to iPATH, and how particles are accelerated at the shock front,  trapped behind the shock front in the shells in the iPATH model. 
Section~\ref{sec:results}  contains the analyses for the two events observed by SolO. The observed upstream wave intensities are used to obtain the turbulence decay time scales for both events. These time scales are used to drive the effective length scale of enhanced upstream turbulence in the iPATH model.  The observed duration of the enhanced downstream turbulence was compared with the shell width from the simulations. We also carefully discuss the escape process upstream of the shock. For a particular energy-dependent choice of the escape length, qualitative agreements between observation and simulation can be obtained.   The main conclusions of this work are summarized in Section~\ref{sec:conclu}.

\section{Model Setup}\label{sec:model}
\subsection{Modelling CME and its driven shock}

EUHFORIA is a comprehensive data-driven coronal and heliospheric model specifically designed for space weather forecasting, combining two major modules to simulate the realistic solar wind conditions of the inner heliosphere: the empirical coronal model and the heliospheric MHD model \citep{pomoell2018euhforia}.  In this study, the CME  is simulated using the Cone model \citep{zhao2002determination,odstrcil2004numerical}. The cone model simulates the CME as a hydrodynamic cloud of plasma with increased density and temperature. It is inserted into the solar wind with a constant speed and angular width. We adopt CME parameters in the Space Weather Database Of Notifications, Knowledge, Information \footnote{\url{https://kauai.ccmc.gsfc.nasa.gov/DONKI/}} (DONKI ) and in the CDAW catalog\footnote{\url{https://cdaw.gsfc.nasa.gov/CME_list/HALO/halo.html}} as references. The in-situ plasma and magnetic field measurements provide information on the arrival of the shock and the solar wind conditions upstream and downstream of the shock. The kinematic insertion parameters (the CME speed and density) are fine-tuned to match the shock arrival time and in situ plasma measurements at SolO. The specific parameters for the insertion of the cone CME in two events are provided in Table~\ref{table1-Cone}. Grid resolutions in EUHFORIA are as follows: 1024 grid cells in the radial direction between 0.1 au and 2.0 au, and a $4^{\circ}$ angular resolution in longitudes and latitudes. 

In the context of studying CME-driven shocks in EUHFORIA, it is essential to accurately identify the shock structure. To achieve this, we adopt the methodology described in \cite{ding2022A&A...668A..71D}. The initial step is the identification of shock positions within the simulation. Once the shock locations have been determined, we calculate several important shock parameters, including the shock speed, shock compression ratio, and shock obliquity.  The shock locations and the shock parameters serve as crucial inputs for the iPATH model, which considers the dynamics variation of CME-driven shocks and flow conditions upstream and downstream of the shock.

\begin{table*}
\caption{\label{table1-Cone}Input parameters of the Cone CME model in the EUHFORIA}
\centering
\begin{tabular}{lccc}
\hline\hline
Parameter & Event 1 & Event 2\\
\hline
Insertion time    &  2022-08-30T20:29:00       &  2022-09-05T18:24:00 \\
Insertion latitude (HEEQ)   & -15$^\circ$    & -25$^\circ$  \\
Insertion longitude (HEEQ) &  150$^\circ$ &  175$^\circ$ \\
half-width        &  50 $^\circ$    &  40 $^\circ$ \\
Speed           &  1000 km/s      &  2200 km/s \\
Density    &   $4.0$ $\times$ $10^{-18}$ kg m$^{-3}$  &   $0.5$ $\times$ $10^{-18}$ kg m$^{-3}$\\
Temperature    &  $2.0$ $\times$ $10^6$ K   &  $2.0$ $\times$ $10^6$ K \\
\hline
\end{tabular}
\end{table*}

\subsection{The iPATH model}
A detailed discussion of the iPATH model can be found in  \citet{Hu+etal+2017}. Here we briefly discuss the structures of the iPATH model and the relevant parameters in iPATH model that are relevant to our current work.
The iPATH model contains three modules: (1) an MHD module that simulates the background solar wind and the CME-driven shock and tracks the downstream shell structures; (2) a particle acceleration module that computes particle spectra at the shock front; and (3) a particle transport module follows the propagation of particles escaping upstream of the CME-driven shock. In this work we use EUHFORIA to replace the MHD module of the iPATH code. The coupling of EUHFORIA and iPATH models has been introduced in \cite{ding2022A&A...668A..71D}, which is similar to the approach taken in \cite{Li2021}  where the authors coupled the Alfv\'en Wave Solar Model (AWSoM; \citet{vanderholst10}) with the iPATH to provide a more realistic description of the CME-driven shock.  

Following \citet{Ding+2020,Li2021,ding2022A&A...668A..71D}, we describe the instantaneous particle distribution function at the shock by,
\begin{equation}
f(\mathbf{r}, p,t_k) = c_1\epsilon_{\mathbf{r}}n_{\mathbf{r}}p^{-\beta}H[p-p_{\rm inj, \mathbf{r}}] 
\exp\left[-\left(\frac{E}{E_{0,\mathbf{r}}}\right)^{\alpha}\right],
\label{eq:fp}
\end{equation}
where $\beta = 3s_{\bf{r}}/s_{\bf{r}}-1$, $s_{\bf{r}}$ is the shock compression ratio at $\bf{r}$(r,$\theta$,$\phi$), $\epsilon_{\bf{r}}$ is the injection efficiency, $n_{\bf{r}}$ is the upstream solar wind density,
$p_{\rm inj,\bf{r}}$ is the particle injection momentum, $E_{0,\mathbf{r}}$ is the kinetic energy that corresponds to a maximum proton momentum $p_{\rm max,\bf{r}}$. The exponential tail  $\rm exp(-(E/E_0)^{\alpha})$ accounts for
the finite shock extension and finite acceleration time 
and $\alpha$ is a free parameter to describe the steepness of the exponential tail. We adopt $\alpha=2$ as \cite{ding2022A&A...668A..71D}. The injection rate is assumed to be $0.5\%$ at the parallel shock. $H$ is the Heaviside function, and $c_1$ is a normalization constant given by
\begin{equation}
c_1 = 1/\int_{p_{\rm inj,\bf{r}}}^{+\infty}p^{-\beta} H[p-p_{\rm inj,\bf{r}}]*\exp\left[-\left(\frac{E}{E_{0,\mathbf{r}}}\right)^{\alpha}\right]d^3p.
\end{equation}

In the 2D iPATH model \citep{Hu+etal+2017} the accelerated particles that convect with the shock and diffuse downstream of the shock are tracked using a shell model. A 3D version of the shell model has been developed with the data-driven MHD models \citep{Li2021,ding2022A&A...668A..71D} where individual shells are divided into multiple parcels that are labelled by their longitudes and latitudes.  The angular resolution is $4$ degrees, the grid resolution of EUHFORIA. Following \citet{ding2022A&A...668A..71D},  we only consider the evolution of the shell along the radial direction. For a given ($\theta,\phi$), the outer edge of the outermost shell is given by the shock front $r_i$ ($i$ is the number of time steps) at time $t_i$.  Other shells with the same ($\theta,\phi$) have their outer edges at radial distances $r_j$ (j = 1,2,...,i-1), which are functions of time. At  time $t_i$,   $r_j$  are given by,
\begin{equation}\label{eq:shell_1}
r_{j}(t_{i}) = r_{j}(t_{i-1})+\int_{t_{i-1}}^{t_{i}}u(r_{j}(t_{i-1}+t'),\theta,\phi)dt',
\end{equation}
where $u$ is the solar wind speed at the shell location ($r_j,\theta,\phi$) at successive MHD time steps. In discrete form, Eq.~(\ref{eq:shell_1}) becomes,
\begin{equation}\label{eq:shell2}
r_{j}(t_{i}) = r_{j}(t_{i-1})+ u(r_{j}(t_{i-1}),\theta,\phi)(t_{i}-t_{i-1}).
\end{equation}
Equation~(\ref{eq:shell2}) enables us to construct all parcels within the shell using the outputs of the EUHFORIA model. The shell model in iPATH is constructed  via the realistic 3D shock fronts and time-dependent downstream flow speed from the underlying MHD code. It therefore captures the spatial extension of the downstream region of the shock.  For different events the shell extension can differ significantly. Note that as an implicit assumption in the iPATH model, the turbulence is assumed to be much enhanced in the shells. This implies that a realistic shell model can be used to understand the duration of the shock sheath, where enhanced turbulence is often found. 
We also note that the duration of the shock sheath can serve as a good approximation of the duration of the ESP phase. This is because that if the CME is composed by closed field lines, then to first approximation,  particles accelerated at the shock front, presumably along open field lines, can not 
penetrated into the CME ejecta.  However, if the ejecta contains a significant amount of open field lines, SEPs can indeed penetrate in to CME ejectra. Depending on how many open field lines are present, a decrease of SEP intensity is expected. Because different shells are constructed when the shock is at different times, the resulting energetic particle distributions in these shells are therefore different. As the shock propagates out in time, these different energetic particle populations will be mixed since these particles can diffuse and convect among different parcels 
\citep{Hu+etal+2017,Ding+2020}. This mixing leads to a time-dependent downstream energetic particle distribution, which can be compared with observations.  

When particles diffuse far enough upstream of the shock, they can escape.  We discuss the process of particle escape upstream of the shock in Section\ref{subsec:escape }. In the iPATH model, the transport of these particles in the solar wind is described  by a focused transport equation. We follow \cite{ding2022A&A...668A..71D} in modelling this transport process.

\section{Results}\label{sec:results}
\subsection{Solar Orbiter Observations}\label{subsec:solo observation}

\begin{figure*}
\includegraphics[width=18.9cm]{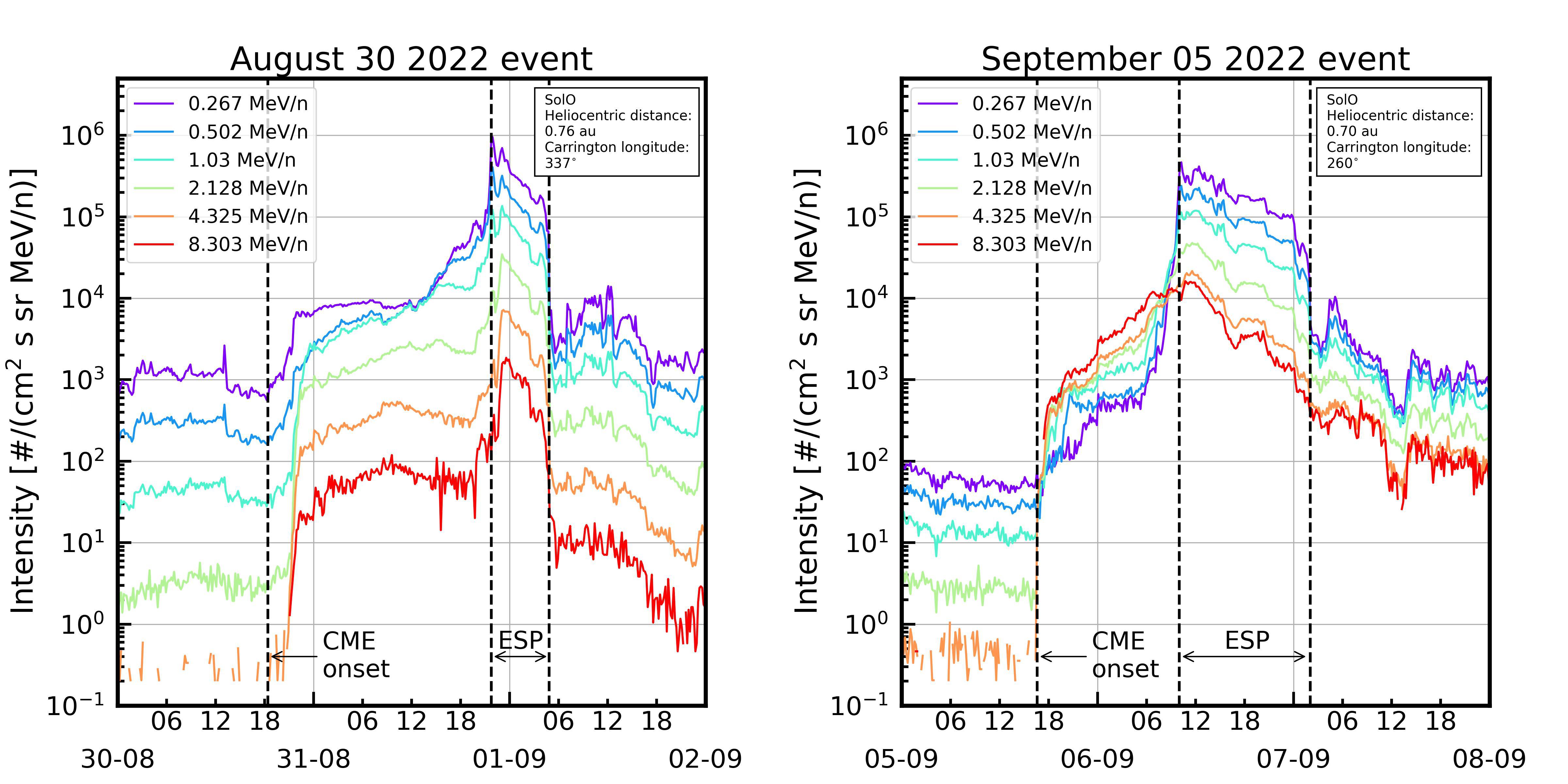}
\caption{Proton time intensity profiles  for the August 30 2022 (left) and the September 05 2022 (right) events as observed by SolO/SIS. The energy range is  from $0.267$ MeV/n to $8.303$ MeV/n. The left, middle and right dashed lines in both panels indicate the onset time of CME, the arrival time of the IP shock and the end of the ESP event at SolO, respectively. }
\label{fig1:time intensity profile}
\end{figure*}

We utilize in-situ measurements of solar energetic particles, plasma, and magnetic fields obtained by the Suprathermal Ion Spectrograph (SIS) within the Energetic Particle Detector (EPD) suite \citep{Pacheco2020A&A...642A...7R,Wimmer-Schweingruber2021A&A...656A..22W}, Solar Wind Analyser \citep{Owen2020A&A...642A..16O}, and MAG magnetometer \citep{Horbury2020A&A...642A...9H} onboard the Solar Orbiter. These measurements are available most of the time between August 30 2022 and September 8 2022, providing an excellent opportunity for comprehensive analysis and comparison of the two ESP events. In this study, our focus is on analyzing the proton intensity measurements obtained by the EPD/SIS instrument and comparing them with the simulated results generated by the combined EUHFORIA and iPATH models. The recent measurements conducted by the EPD/SIS instrument have provided valuable insights into the energetic and suprathermal ion composition in various energetic particle events, including quiet-time ion composition \citep{Mason2021A&A...656L...1M,Mason2023A&A...673L..12M}, large and small SEP events \citep{Ho2022FrASS...9.9799H,Bucik2023A&A...673L...5B,Mason2021A&Asep,Mason2021A&A3He-rich}, as well as corotating interaction region (CIR) events \citep{Allen2021A&A...656L...2A}. Some recent unexpected observations of SEP and SIR-related ion events by the EPD are summarized in the review of \cite{Malandraki2023PhPl...30e0501M}.  The EPD/SIS sensor has demonstrated exceptional sensitivity, enabling precise measurements of the intensity of low-energy channels.

Figure~\ref{fig1:time intensity profile} shows the 10-min average proton time intensity profiles observed by SolO/SIS for the August 30 and the September 5 events. The plot shows the average of the SIS sunward and anti-sunward telescopes. Six energy channels ranging from $0.267$ to $8.303$ MeV/n are selected for analysis. Several notable characteristics emerge when comparing these two events. Firstly, the durations of the ESP events differ significantly, with the August event lasting for approximately $7$ hours and the September event extending over a period of $16$ hours. These durations correspond to the passage of the shock-sheath structure associated with each event. The duration of particles being trapped upstream of the shock is difficult to determine precisely, which is energy-dependent and affected by the transport effects. We note that SolO is located at similar solar distances, $0.76$ au for the August event and $0.7$ au for the September event.    Secondly, in the September event it can be seen that near the shock the intensities of low-energy protons are lower than that of high-energy protons, resulting in a unique ``cross-over'' feature in the time profiles prior to the arrival of the shock. Note that this ``cross-over'' is different from the phenomenon related with velocity dispersion seen near the onset of the SEP events (e.g., \cite{Mason+2012, wu_statistical2023}). There,  both lower and higher energy particles injected close to the Sun at the same  time and higher energy particles arrive at observer first. The ``cross-over'' in our event occurs close to the shock, and as we discuss below, it is caused by a long-lasting turbulence-enhancement upstream medium. 
Such a behavior is rare in gradual SEP events, and to our knowledge, only one similar event was reported by \citet{lario2021comparative}, who analyzed 
the November 29 2020 event observed by the PSP. They suggested that such a time profile could be related to pre-existing interplanetary coronal mass ejection (ICME) structures. As we discuss below, we believe this ``cross-over'' is a feature of particle escape upstream a CME-driven shock, and its infrequent appearance is related to the escape length of particles, which is in general a function of particle energy and which varies from event to event.

\begin{figure*}
\includegraphics[width=18.9cm]{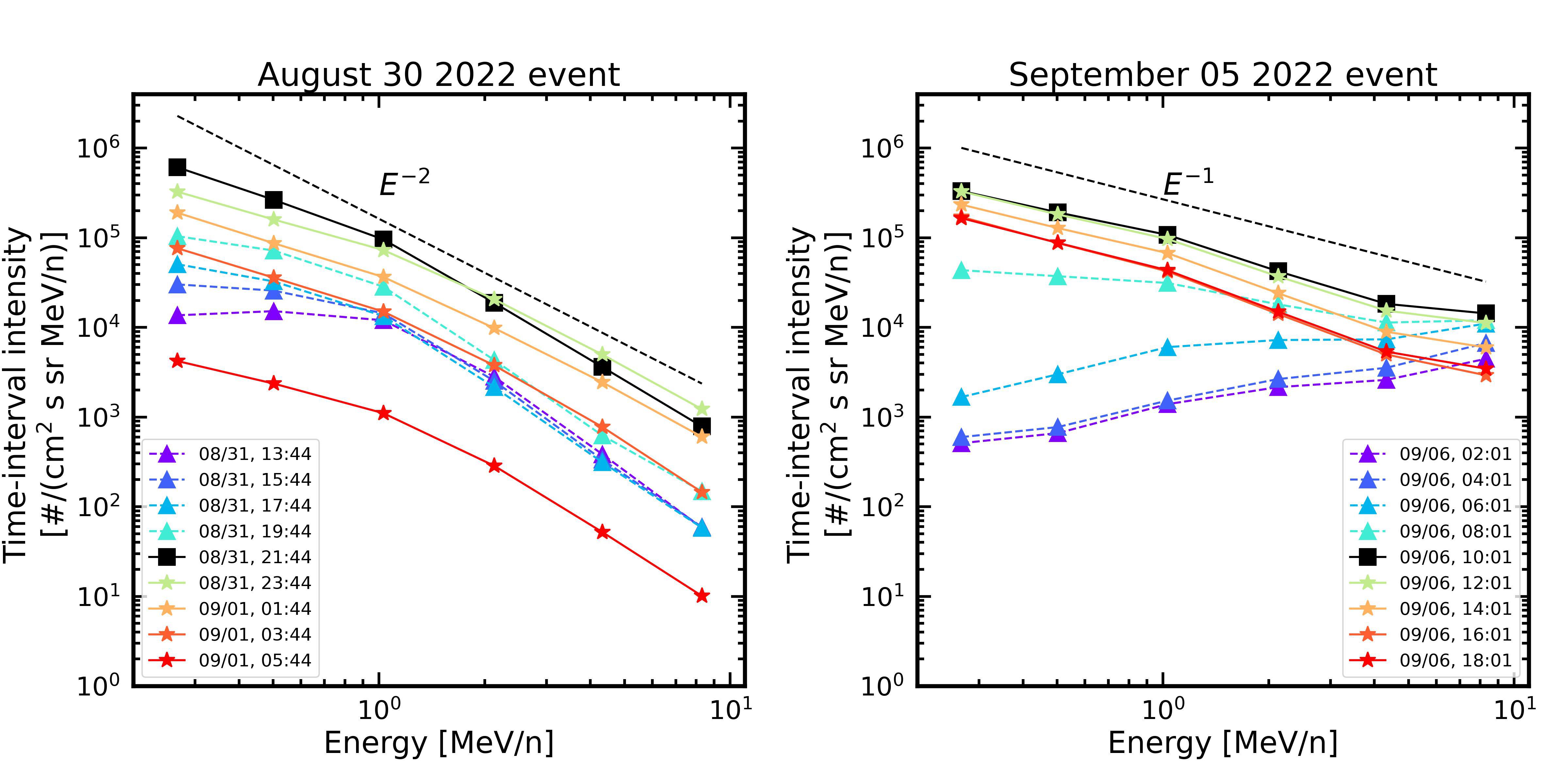}
\caption{2-hour time-interval proton spectra for the August 30 2022 (left) and the September 05 2022 (right) events as observed by SolO/SIS from $0.267$ MeV/n to $8.303$ MeV/n. The timestamps represent the start of the intervals, ranging from $8$ hours before to $8$ hours after the arrival of the shock. The black curve with squares corresponds to the time of shock arrival. Triangles and stars represent observations upstream and downstream of the shock. The dashed black line is to guide the eye. It is $E^{-2}$ for the August event and $E^{-1}$ for the September event. Upstream the September 05 event, the ``cross-over'' is clearly seen.}
\label{fig2:SIS spectra}
\end{figure*}

Figure~\ref{fig2:SIS spectra} presents 2-hour time-interval proton spectra for the August 30 2022 (left), and the September 05 2022 event  (right), as observed by SolO/SIS in the energy range from $0.267$ MeV/n to $8.303$ MeV/n. We focus on the spectra within an 8-hour window observed upstream and downstream of the shock. During this period, the August event shows mostly a power law with an index of $\sim -2$, which is normal for SEP events \citep{ebert2016ApJ...831..153E,Desai2016ions,Duenas2022}. Upstream of the shock, there is a slight bend-over feature observed for energies below $1$ MeV, which is commonly attributed to a transport modulation effect. Compared to the August event, the September event exhibits a distinctly different behavior where the spectra farther out from the shock are positive-power-law like, and it transitions to a normal negative-power-law like spectrum across the shock. Additionally, the spectra downstream of the shock are harder compared to the August event, with a power law index of approximately $-1$. The presence of a positive power law in the spectra of the September event, extending up to energies of  $8.3$ MeV/n, makes it challenging to explain solely based on the transport effects in the solar wind.

\begin{figure*}
\includegraphics[width=18.9cm]{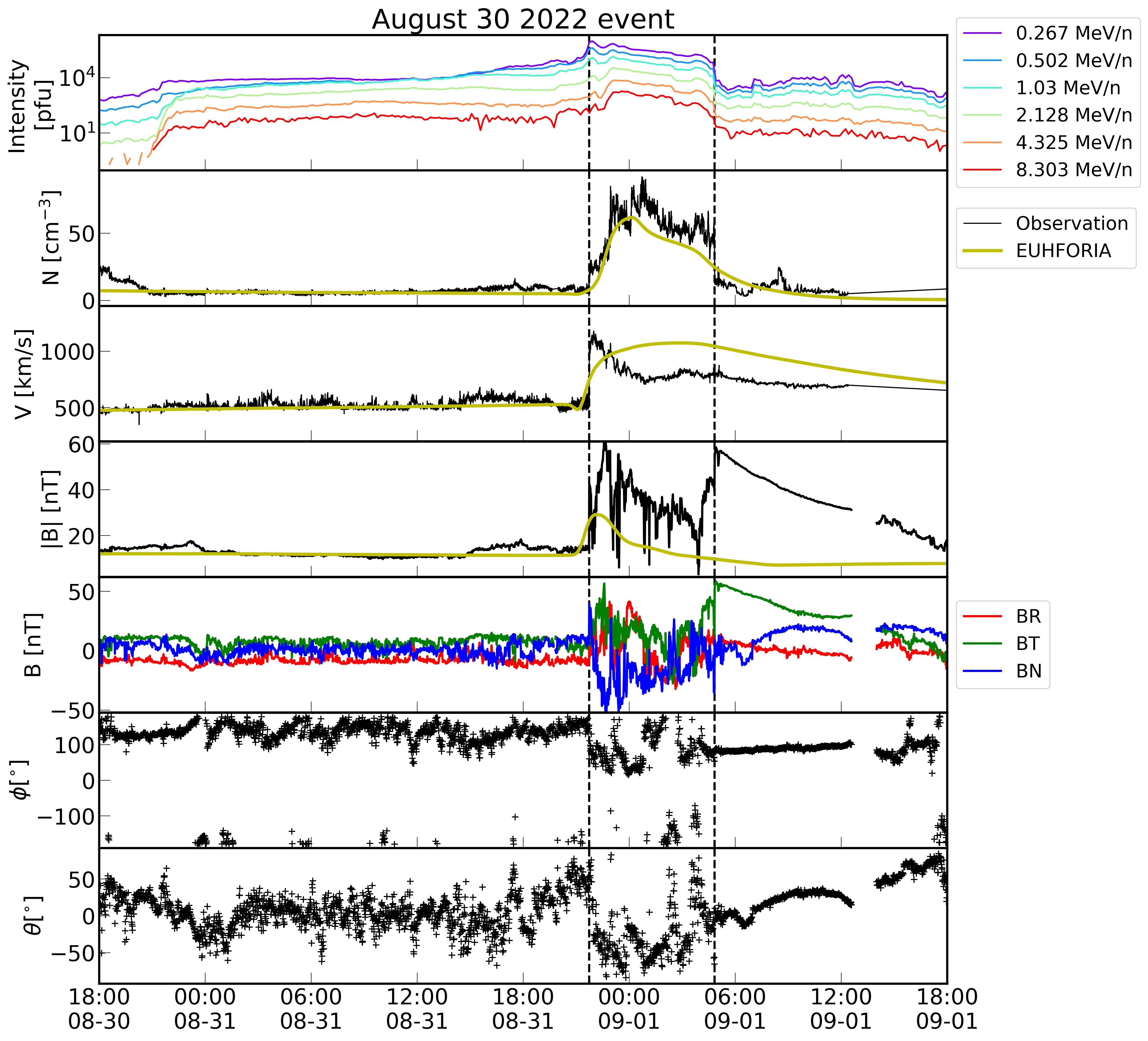}
\caption{SEP time profiles and in-situ plasma and magnetic field for the August 30 2022 event. The panels present, from top to bottom, the energetic proton time profiles, the solar wind  proton number density, the solar wind speed, the magnetic field magnitude, the magnetic field vector measured in the RTN coordinate system, azimuthal ($\phi$) and elevation ($\theta$) angles of the unit vector magnetic field. The yellow lines are the EUHFORIA simulation results. The dashed lines indicate the IP shock and the end of the ESP event.}
\label{fig3:insitu measurement-August}
\end{figure*}

We further examine the plasma and magnetic fields measured by SolO for the two events. Figure~\ref{fig3:insitu measurement-August} shows the measurements  for the August event. From top to bottom, these are energetic proton time profiles, solar wind proton number density, the solar wind speed, the magnetic field magnitude, the magnetic field vector measured in the RTN coordinate system, azimuthal ($\phi$) and elevation ($\theta$) angles of the magnetic direction. In addition to the observed data, we also include the simulated solar wind density, solar wind speed and magnetic field magnitude from the EUHFORIA model, represented by the solid yellow lines in Fig.~\ref{fig3:insitu measurement-August}. Further details of the EUHFORIA simulations are discussed in Section~\ref{subsec:shell_euhforia }. As shown in Fig.~\ref{fig3:insitu measurement-August}, before the arrival of the shock, both the plasma and magnetic field conditions are relatively undisturbed. After the shock, a distinct sheath structure and a magnetic cloud are observed. The duration of the ESP event is clearly bounded by the sheath structure, lasting  $\sim 7$ hours. Furthermore, it is worth noting that the peak intensity of high-energy protons in this event occurs within the sheath region, approximately $1$ hr after the shock passage.  Two similar events with proton intensity peaks after the shock passage were reported in \cite{Giacalone2012ApJ...761...28G}. This differs from typical ESP events where the peak intensity is usually observed at the location of the shock \citep[e.g., ][]{Lario2018JPhCS1100a2014L,Lario2023ApJ...950...89L}. Understanding and simulating this particular feature is beyond the scope of this work.

\begin{figure*}
\includegraphics[width=18.9cm]{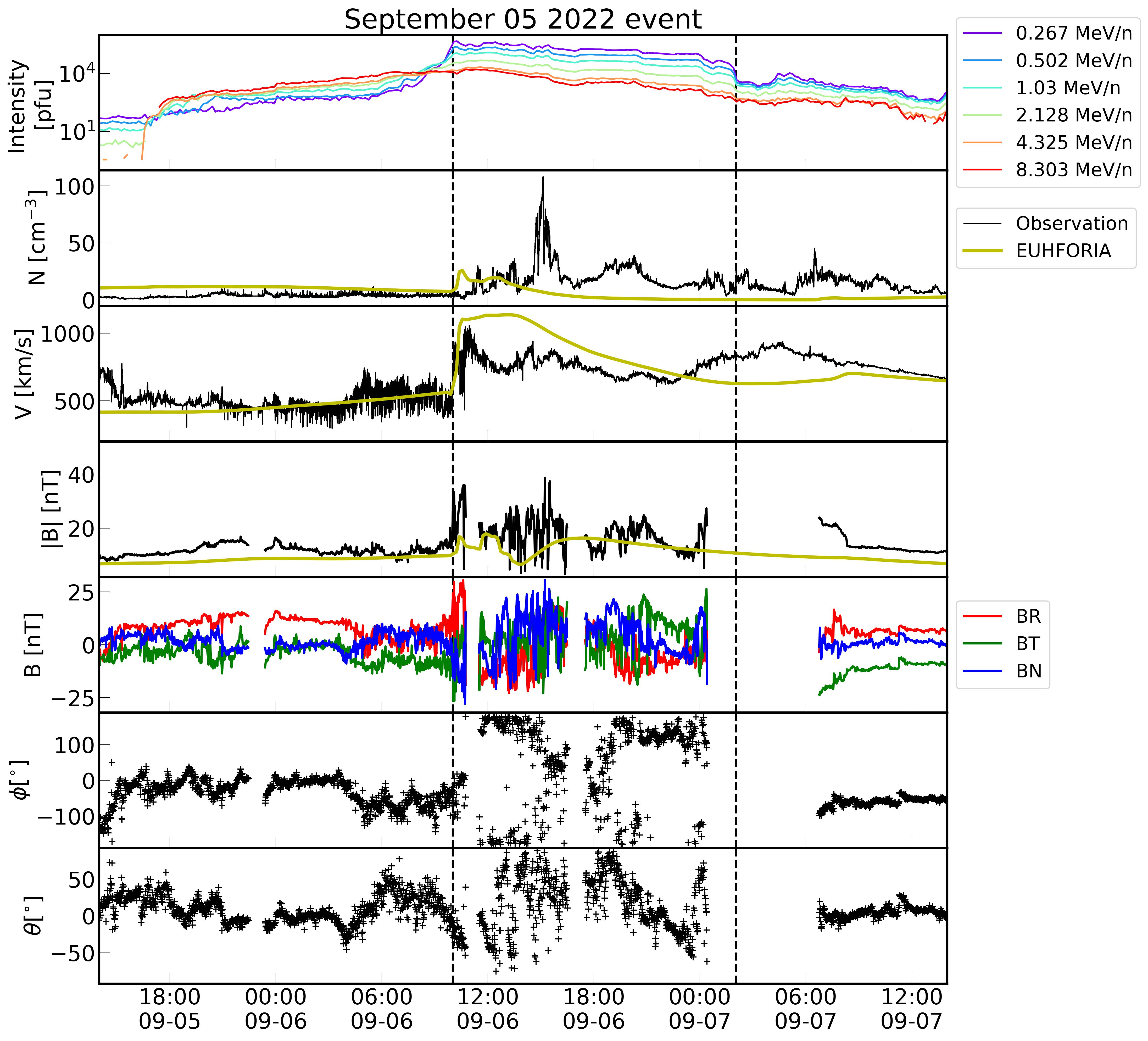}
\caption{Same as Fig.~\ref{fig3:insitu measurement-August}, but for the September 05 2022 event.}
\label{fig4:insitu measurement-September}
\end{figure*}

The measurements of plasma and magnetic fields for the September event are shown in Figure~\ref{fig4:insitu measurement-September}.  A comparison with the August event reveals more enhanced fluctuations in the flow velocity and magnetic field upstream of the shock. In particular, the azimuthal ($\phi$) and the elevation ($\theta$) angles of magnetic direction vary significantly for about $6$ hours ahead of the shock. 
Upstream from the shock, the intensities of low energy particles (e.g., $1.03$ MeV and below) are smaller than those of higher energies (e.g., 4 MeV and 8 MeV). \citet{lario2021comparative} reported a similar ``cross-over''  ESP event on 2020 November 29. They suggested that the low-energy particles are excluded by an ICME upstream of the shock. 
However, in that event, the ``cross-over'' feature did not recover back to normal time profiles after the  ICME passed through PSP and was preserved until the shock arrived at the observer. They also compared several historical ESP events with proceeding ICMEs but those events did not exhibit the ``cross-over'' feature. Therefore, it is the opinion of the authors of this article that preceding ICMEs are not the cause for the observed ``cross-over'' feature. 
Indeed, there was no clear magnetic cloud detected upstream of the shock in the September event. We note that, however, there are intense magnetic fluctuations upstream of the shock, indicating the presence of strong turbulence ahead of the shock.

\begin{figure*}
\includegraphics[width=18.9cm]{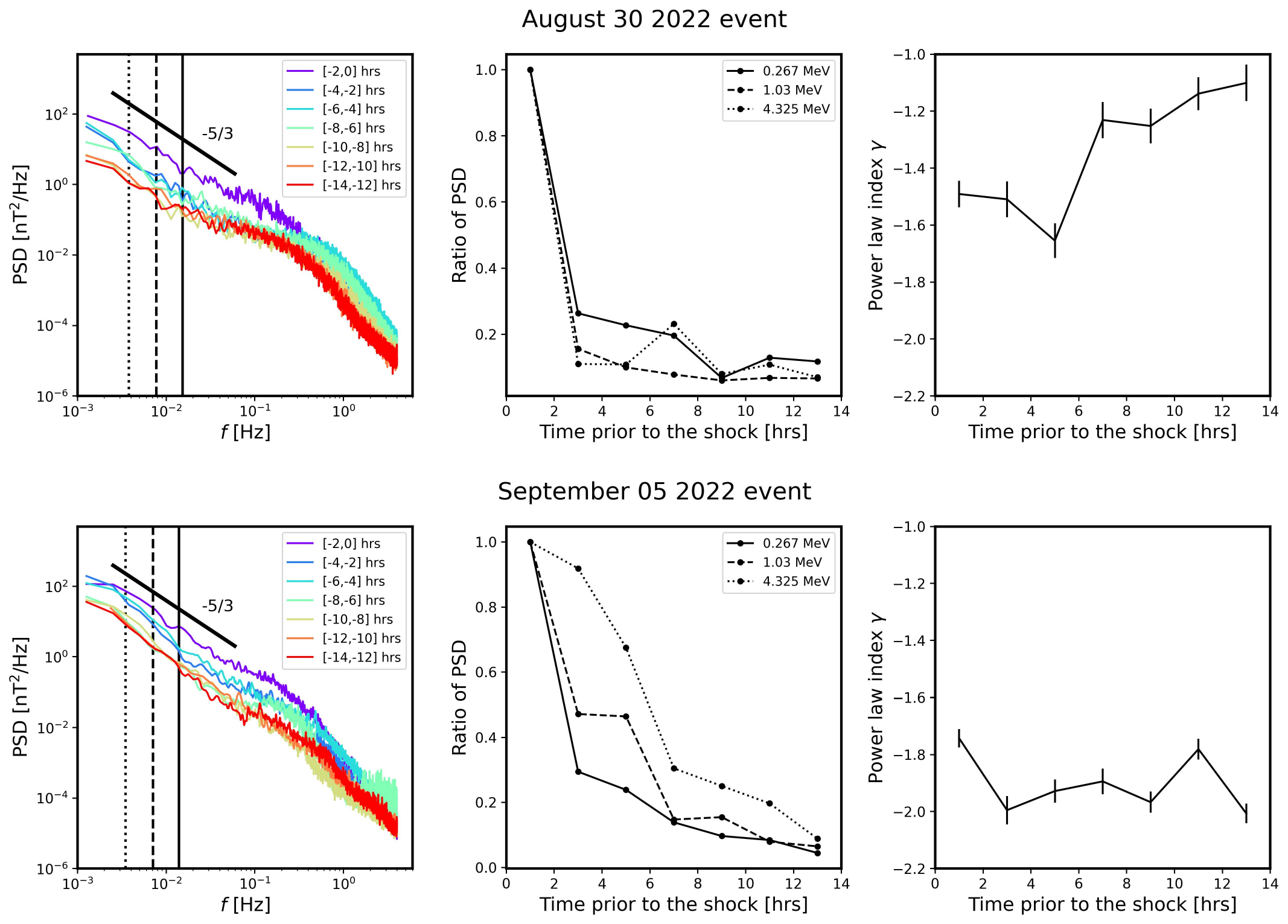}
\caption{Total magnetic field Power Spectral Density (PSD) for August 30 2022 (upper) and September 05 2022 (lower) events. The left panels represent the 2-hour interval of PSD as a function of frequency upstream of the shock. The durations are labeled on the top right of the panel. $0$ represents the time of shock. The thick solid line shows the power law of $f^{-5/3}$  for reference. The dashed vertical lines indicate the corresponding resonant frequencies of different energies by Taylor’s hypothesis. The three energies are labelled in the middle panel.   The middle panels show the ratio of PSD  as a function of time upstream of the shock, compared to the PSD closest to the shock (i.e., the PSD for the duration of [$-2$,$0$] hours). The right panels show the fitted power law indices in the frequency range [$0.003$,$0.06$] Hz as a function of time.  }
\label{fig5: power spectral density}
\end{figure*}

To further understand the upstream magnetic fluctuations, we compute the power spectral density (PSD) of the total magnetic field  using Welch's method \citep{welch1967use}.
The power spectra for both events are calculated with a 2-hour interval prior to the shock arrival as shown in Fig.~\ref{fig5: power spectral density}. The left panels show seven power spectra with a 2-hour interval upstream of the shock. 
Energetic particles resonate with these waves and the resonance condition is given by the Doppler relation:
\begin{equation}
\omega - n \Omega - k_{||} \mu v =0;
\label{eq:resonance}
\end{equation}
where $\omega$ is the resonant wave frequency, $n$ is an integer, $\Omega = (Q/A)eB/\gamma m_p c$ is the local ion gyrofrequency, $k_{||}$ is the wave vector component along the
background magnetic field, and $\mu$ is the particle's pitch angle cosine. In the expression of ion gyrofrequency, $v$ is particle speed in the plasma frame upstream the shock, $Q$ and $A$ are ion charge and mass number, $p$ is particle momentum, $m_p$ is proton mass, $\gamma$ is the Lorentz factor
and $c$ is the speed of light. Since $\omega = k V_A$ for Alfv\'{e}n waves and $V_A$ is a lot smaller than 
particle speed $v$, then under the extreme resonance broadening condition where $\mu \sim 1$ in equation
~(\ref{eq:resonance}), we can obtain the resonance wave number,
\begin{equation}
    k \approx \frac{Q}{A} \frac{eB}{\gamma \beta_v m_p c^2}
\end{equation}
where $\beta_v =v/c$ is particle speed in the unit of $c$.
Using  Taylor's hypothesis \citep{taylor1935statistical}, i.e., $f=kV_{sw}/2\pi$, 
the corresponding resonant frequencies that resonate with $0.267$, $1.03$, and $4.325$ MeV protons are marked by vertical lines (from right to left). Away from the shock, the PSD at these frequencies decays significantly \citep{Hu2013AIPC.1539..175H}. To clarify the decay rate of the PSD at different frequencies (resonating with different particle energies), the ratio of these PSDs, relative to that closest to the shock (the duration of [-2,0] hours),  are plotted versus the corresponding time preceding the shock passage in the middle panels. 
The decay rates for these two events differ significantly. For instance, the ratio decreases to $0.1$ rapidly after $\sim 2$ hours for the August event, while it takes $\sim 6$ hours to decay to  $0.1$ for the September event. We then perform a power-law fitting on the spectra between $0.003$ Hz to $0.06$ Hz. Above $0.06$ Hz, the spectra develop a bump-like feature so that we only fit the power spectra between $0.003$ Hz to $0.06$ Hz. 
The right panel plots the spectral index as a function of time.  We find a significant difference in the spectral slopes between the two events. The September event exhibits a steeper spectral slope of approximately $f^{-2}$ and does not vary for 12 hours upstream of the shock, suggesting a well-developed and strong turbulence ahead of the shock \citep{Li+2003}.

\subsection{Particle escape upstream of the shock }\label{subsec:escape }

\begin{figure*}
\includegraphics[width=18.9cm]{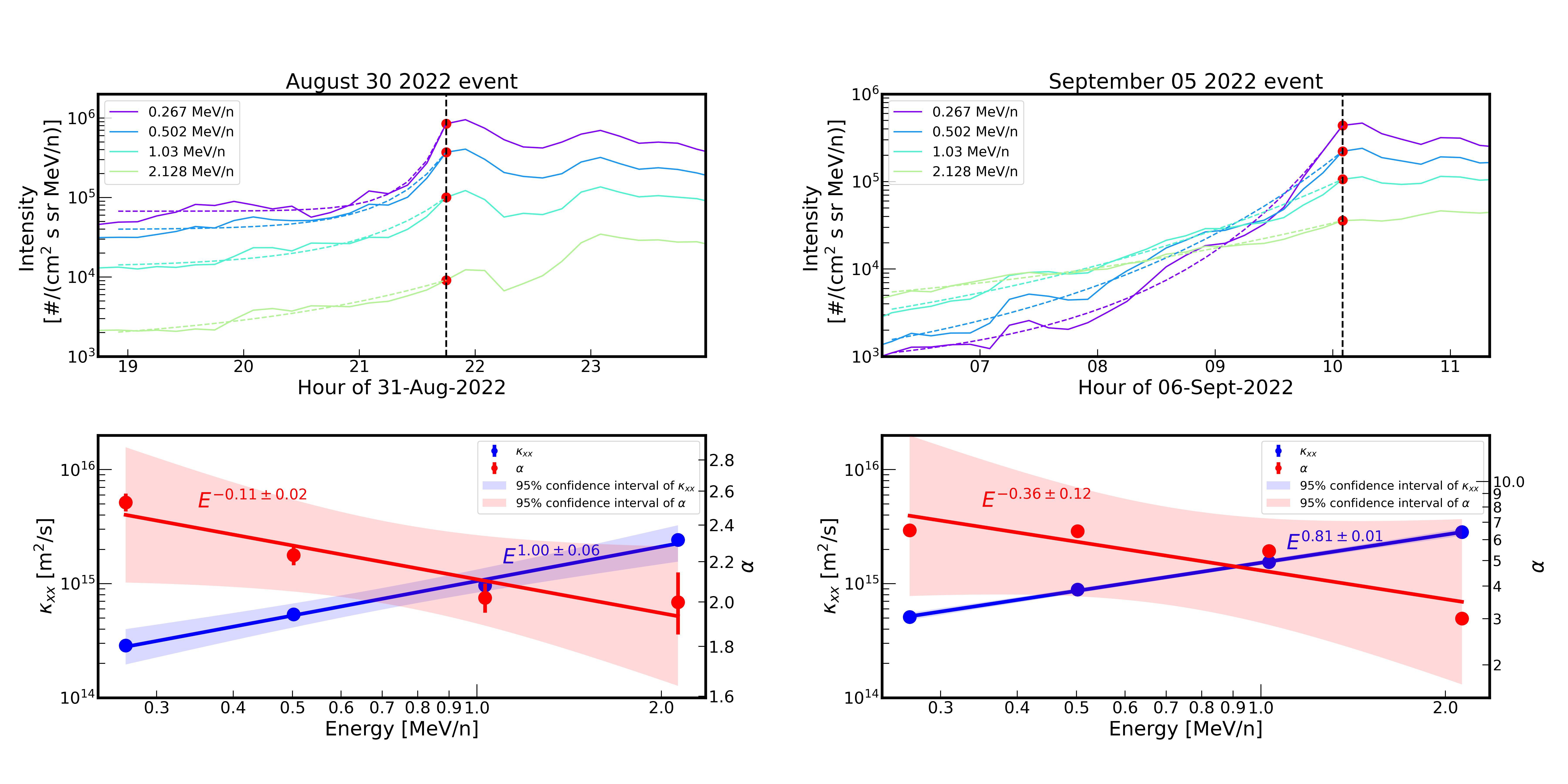}
\caption{ Upper panels: The solid (dashed) lines represent the observations (fits) of time profiles upstream of the shock by equation~(\ref{eq:1d_fp_solution_exp_waveintensity}). Lower panel: Diffusion coefficients and the index $\alpha$ as a function of energy for the two events. The lines show  the power law fits of the two parameters with $95$\% confidence interval. See text for details.  
}
\label{fig6:fitted kappa}
\end{figure*}

As we demonstrate below, the very different behavior for these two ESP events can be attributed to very different upstream turbulence environments which impact the escape process in these two events. To understand the observations, we examine the escape process in more detail in this section.

We assume the particle acceleration is due to the first-order Fermi acceleration and the particle scattering is isotropic in the shock frame. At the shock front, the particle distribution function $f$ satisfies the Parker transport equation:
\begin{equation}\label{eq:transport equation}
 \frac{\partial  f}{ \partial t} + \mathbf{U} \cdot \nabla f = \nabla \cdot (\mathbf{\kappa} \nabla f)+ \frac{1}{3} \nabla \cdot \mathbf{U} \frac{\partial f}{ \partial \ln p} + S - L
\end{equation}
where $\mathbf{U}$ the plasma bulk velocity, $\mathbf{\kappa}$ the spatial diffusion tensor, and $p$  the particle momentum. The second term on the left represents the convection and the terms on the right represent the spatial diffusion, energy change, sources,  and losses respectively. 

Considering a one-dimensional planar shock with an injection momentum $p_0$, the steady-state solution of $f$ is given by \cite[e.g.,][]{Drury+1983}
\begin{equation}\label{eq:1d_fp_solution}
f\left(x,p\right)= C p^{-3s/\left(s-1\right)}H\left(p-p_{0}\right) \exp \int^{x}_{0} \frac{U_1}{\kappa_{xx}(x',p)}dx'
\end{equation}
where $C$ is a constant, $s$ is the compression ratio, $x$ is the distance to the shock which is at $x=0$ ($x<0$ represents upstream and $x>0$ represents downstream  ), $U_1$ is the upstream speed in the shock frame, $\kappa_{xx}(x,p)$ is the components of $\mathbf{\kappa}$ along the shock normal direction as a function of $x$ and $p$, and $H$ is the Heaviside step function. Following \cite{Li+2005ions},  the diffusion coefficient $\kappa_{xx}(x,p)$ is proportional to the wave intensity inverse $I(k_{res},x)^{-1}$ evaluated at the resonant wave number $k_{res}$ as given by equation~(\ref{eq:resonance}) (again, assuming the extreme resonance broadening condition).
For an x-independent diffusion coefficient $\Tilde{\kappa}_{xx}$, which is possible if the wave intensity $I(k,x)$ upstream of the shock is x-independent, the solution of the upstream distribution function has the following form,
\begin{equation}\label{eq:1d_fp_solution_L1_Ldiff}
f\left(x,p\right)= C p^{-3s/\left(s-1\right)}H\left(p-p_{0}\right) \exp (\frac{-|x|}{L_{\rm diff}(p) })
\end{equation}
Here, the diffusion length is defined as $L_{\rm diff}(p) = \Tilde{\kappa}_{xx}(p)/ U_1$. Equation~(\ref{eq:1d_fp_solution_L1_Ldiff}) has been examined in many previous studies \citep[e.g.,][]{Zank+etal+2000, Li+2003, Li+2005ions, Wijsen2022A&A...659A.187W}, which signals the exponential decay behavior often seen in SEP observations. Of course, 
the wave intensity upstream the shock is not a constant, and often itself decays with distance.  If we denote $L_{\rm esc}$ as the characteristic length scale describing the decay of upstream wave intensity,  then the behavior of upstream particle distribution depends on the comparison of these two scales. Note that when $I(k,x)$ is x-dependent, one can define $L_{\rm diff}$ using either $I(k,0)$ or some x-averaged $I(k)$. 

The exact decay behavior of wave intensity is unclear and likely event-dependent. As a simplified assumption, we consider an exponential increase of the diffusion coefficient upstream of the shock.  This leads to an estimation of $\kappa_{xx}(x,p) = \kappa_{xx}(x_0,p) \exp (|x|/L_{\rm esc}(p)) $, where $\kappa_{xx}(x_0,p)$ is the diffusion coefficient at the shock ($x_{0}$) and $L_{\rm esc}(p)$ is the length scale of upstream wave intensity.  
Equation~(\ref{eq:1d_fp_solution}) now becomes, 
\begin{equation}\label{eq:1d_fp_solution_exp_waveintensity}
f\left(x,p\right)= C p^{-3s/\left(s-1\right)}H\left(p-p_{0}\right) \exp (-\frac{U_1 L_{\rm esc}(p)}{\kappa_{xx}(x_0,p)} (1 - \exp(-\frac{|x|}{L_{\rm esc}(p)} )).
\end{equation}
Note that when the 
wave intensity decays exponentially away from the shock, the particle intensity upstream of the shock does not decrease to zero when $|x| \to \infty$ as shown in equation~(\ref{eq:1d_fp_solution_L1_Ldiff}). Instead, it approaches a constant when $|x| \gg L_{\rm esc}$, 
indicating that particles can readily escape at some distance from the shock where the turbulence becomes less significant.

The effect of the length scale $L_{\rm esc}$ can be visualized as an escape term $f/\tau$ in the transport equation: particles moving to a distance  $L_{\rm esc}$ upstream of the shock escape from the system. This was done in  \cite{Li2005AIPC_upstreamturbulence} who introduced the escape term $f/\tau$ and relate $\tau$ to an energy-dependent escape length scale $L_{\rm esc}$.  
 \cite{Li2005AIPC_upstreamturbulence} suggested that the particle distribution function upstream of the shock depends on the ratio of $L_{\rm esc}$ to $L_{\rm diff}$. In previous studies \citep{Zank+etal+2000,Rice+2003, Li+2005ions}, $L_{\rm esc}$ is typically assumed to be $2-4$ times larger than $L_{\rm diff}$.
 However, this assumption does not hold for the September event, where low-energy particles barely escape from the shock, resulting in the observed cross-over profiles. This suggests that $L_{\rm esc}$ should be much larger than $L_{\rm diff}$ and the ratio of $L_{\rm esc}/L_{\rm diff}$ should be energy-dependent.

$L_{\rm esc}$ and its momentum dependence can be different for different events. We introduce a parameter $\alpha$ and 
normalize the escape length to the diffusion length by, 
\begin{equation}
L_{\rm esc}(p)= \alpha(p) L_{\rm diff}(p) .  
\label{eq:alpha_normalization}
\end{equation}
 By fitting the upstream time profiles 
using Equation~(\ref{eq:1d_fp_solution_exp_waveintensity}), we can determine the diffusion coefficient $\kappa_{xx}(x_0, p)$ and $\alpha(p)$ as a function of energy. Figure~\ref{fig6:fitted kappa} shows the energy dependence of the diffusion coefficient and the magnitude of $\alpha$ for both events.  The distance $x$ is obtained from measurement in time by assuming a constant shock speed during this period ($x = V_{\text{shock}} \cdot dt$), where $dt$ is the time from the shock passage. We focus on the energies from $0.267$ MeV/n to $2.128$ MeV/n since the data of $4.325$ MeV/n and  $8.303$ MeV/n channels do not yield satisfactory fitting parameters. We find that both events show a similar energy dependence of the diffusion coefficient, indicating a similar particle diffusion behavior near the shock for both events. The values of $\kappa_{xx}$ are also consistent with some earlier studies \citep{Tan1989JGR....94.6552T,Giacalone2012ApJ...761...28G}.  However, the parameter $\alpha$ differs very much for the two events and has very different energy dependence. For the August event, $\alpha$ ranges from $2$ to $3$ and shows a weak energy dependence $\sim E^{-0.11}$, while for the September event, it ranges from $3$ to $7$ and shows a stronger energy dependence $\sim E^{-0.36}$. So the length scale of enhanced turbulence is larger in the September event, which is consistent with the analysis of the power spectral density of magnetic fluctuations as shown in Figure~\ref{fig5: power spectral density}.  Furthermore, since 
$\alpha \sim E^{-0.36}$, from equation~(\ref{eq:1d_fp_solution_exp_waveintensity}) we see that the exponential decay rate for lower energy is faster than that for higher energy.

\begin{figure*}
\includegraphics[width=18.0cm]{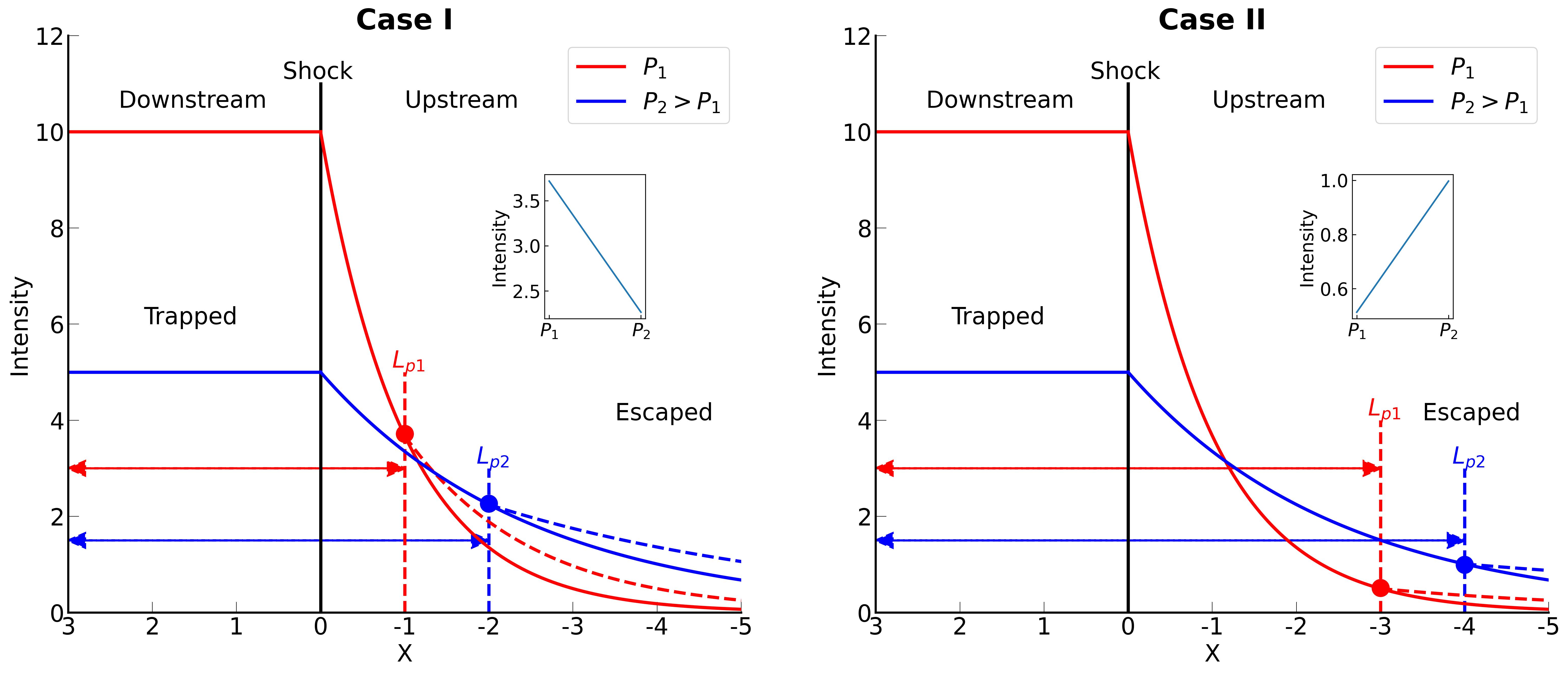}
\caption{Two schematics of the distribution of accelerated particles  at an interplanetary shock for two momenta $P_1$ and $P_2$, adapted from \cite{zank2006}. Both schematics illustrate  the momentum-dependent scale length of the exponential decay upstream of the shock, and the corresponding trapping and escape of particles. The only difference is the different escape length scales $L_{p1}$ and $L_{p2}$. The subpanel shows the intensity of the escaped particles for $P_1$ and $P_2$. The dashed curves ahead of the escape boundary indicate a larger diffusion length scale corresponding to solar wind turbulence. See text for details.  
}
\label{fig7:escape schematic}
\end{figure*}

Figure~\ref{fig7:escape schematic} schematically illustrates the accelerated particle distribution at the shock with different length scales for the enhanced wave intensity $I(k,x)$. 
The steady-state solution of equation~(\ref{eq:transport equation}) shows that the particle intensity is constant downstream of the shock and exponentially decaying upstream of the shock. The exponential decay is determined by the diffusion length scale $L_{\text{diff}}$. $L_{p1}$ and $L_{p2}$ represent the escape boundaries associated with the length scales of enhanced turbulence $L_{\rm esc}$. Particles within the escape boundary are trapped by the self-excited waves, while particles beyond the escape boundary can escape into the ambient solar wind, as indicated by the dashed curves. As shown in Fig.~\ref{fig6:fitted kappa},  the parameter $\alpha$ of escape length scale ($L_{\rm esc}= \alpha L_{\text{diff}}$) for these two events has very different values and energy dependence.  In Case I, where the escape length scale is short, the escaped particle intensity exhibits a negative spectral index, as shown in the subpanel. This is the most common case from observation. However, in Case II, when the escape boundary extends to a greater distance, the spectral index of the escaped particle intensity can invert and become positive. 
Figure~\ref{fig7:escape schematic} schematically explains how a ``cross-over'' and a positive spectral index can occur in the September events. Roughly speaking, the presence of a larger length scale of enhanced turbulence  in the September event hinders the escape of low-energy particles, resulting in the observed ``cross-over'' and the positive spectral index.

\subsection{The duration of ESP event} \label{subsec:shell_euhforia }
\begin{figure*}
\includegraphics[width=18.9cm]{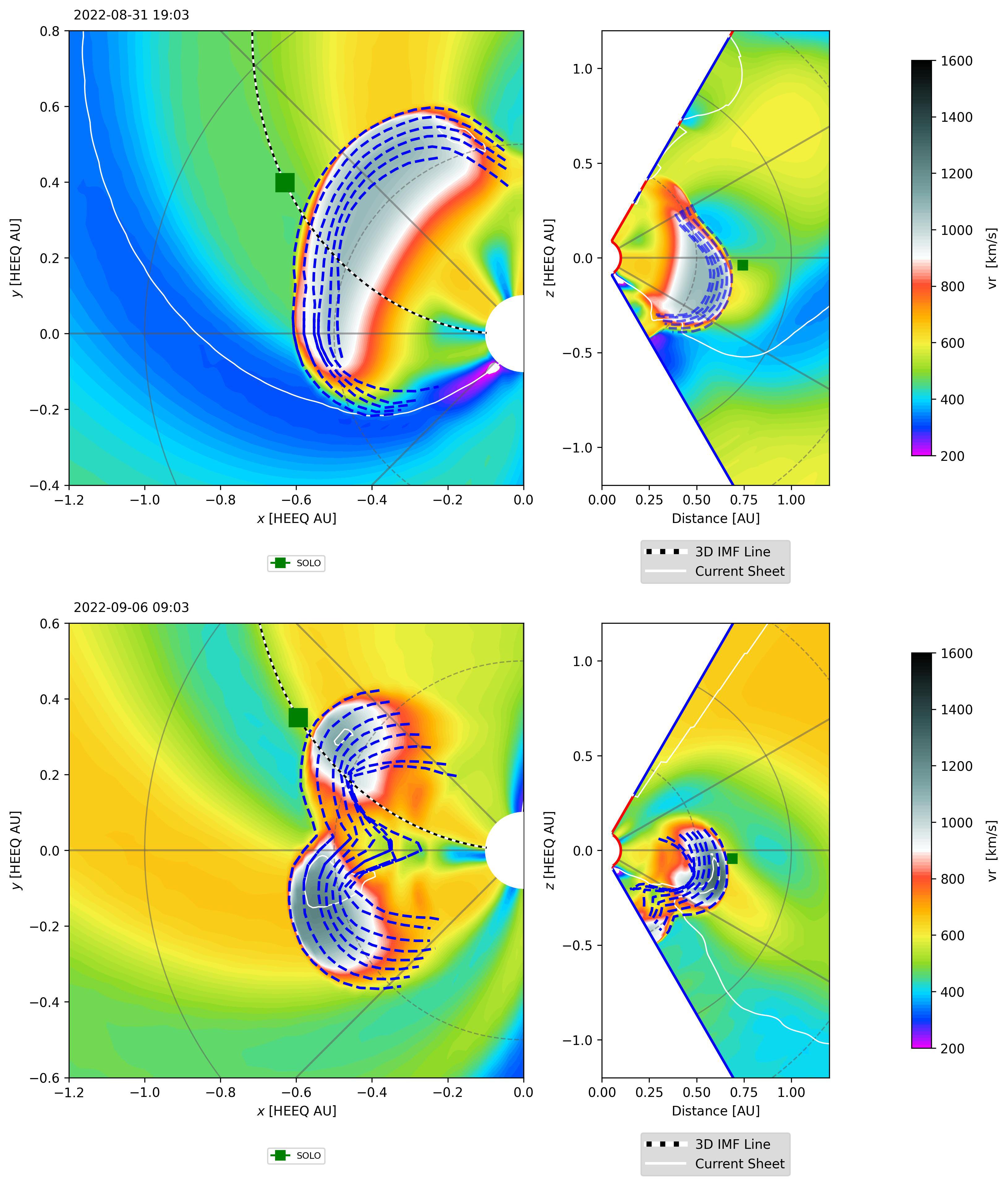}
\caption{ Equatorial (left) and meridional (right) snapshots of the radial solar wind speed from EUHFORIA for the August 30 (upper) and the September 05 (lower) events.  The blue curves show the shell structure behind the shock in the equatorial and meridional planes. }
\label{fig8:euhforia snapshot}
\end{figure*}

Figure~\ref{fig8:euhforia snapshot} shows snapshots of the radial speed of solar wind from the EUHFORIA simulations for the two events, presented in the Heliocentric Earth Equatorial (HEEQ) coordinate system. The corresponding simulated time series of  solar wind parameters are plotted in Figure~\ref{fig3:insitu measurement-August} for the August event and Figure~\ref{fig4:insitu measurement-September} for the September event. For both events, the simulated magnitudes of the number density and the flow speed near the shock are similar to the observations, providing confidence that EUHFORIA simulations can yield a reasonable description of the shell structures that are to be used in the iPATH model. We note that the cone model does not include the magnetic flux rope and therefore cannot accurately account for the magnetic disturbances. As a result, the simulated magnetic field magnitude is not comparable with SolO measurements.
The equatorial and meridional slices of the shell structure are overlaid on top of the CME in Fig.~\ref{fig8:euhforia snapshot}. Comparison between these two events provides insights into the different characteristics of the shell structures. Notably, the August event exhibits a smooth shell front, while the September event displays a two-peak shape. This difference arises from the structure of the shock fronts and the evolution of the downstream speed. The two-peak shape of the shock front is a result of the depression caused by a slow stream with high density ahead of the shock nose. 
Another notable distinction between the events is the radial width of the shell structure. The radial width of the shell structure for the August event is a lot narrower than that of the September event. This difference is attributed to the different deceleration history of the plasma speed downstream of the shock. At the inner boundary,  the two events have very different CME speeds: $1000$ km/s for the August event  and $2200$ km/s for the September event. Observed at SolO,  both events have similar downstream speeds of approximately $1000$ km/s. This implies that there is a strong deceleration downstream of the shock for the  September event. Consequently, the shell structure is stretched more by the rapid deceleration of the downstream flow.  This is particularly evident at the flank of the shock. The deceleration of the shock downstream can also be inferred from the Cone CME parameters. To accurately fit the shock arrival time and the downstream speed observed by SolO, we adopt a small CME density and a small half-width of CME to simulate the pronounced deceleration. 

The radial width of the shell structure is essential for understanding the duration of the ESP event. The arrival of the shock front (the first shell) at the observer marks the onset of the ESP event, while the passage of the last shell marks the end of the ESP event. This means that  the shell module of the iPATH model can serve as a powerful tool to study ESP events. We remark that a data-driven solar wind model is essential for capturing the downstream dynamics.

\begin{figure*}
\includegraphics[width=18.9cm]{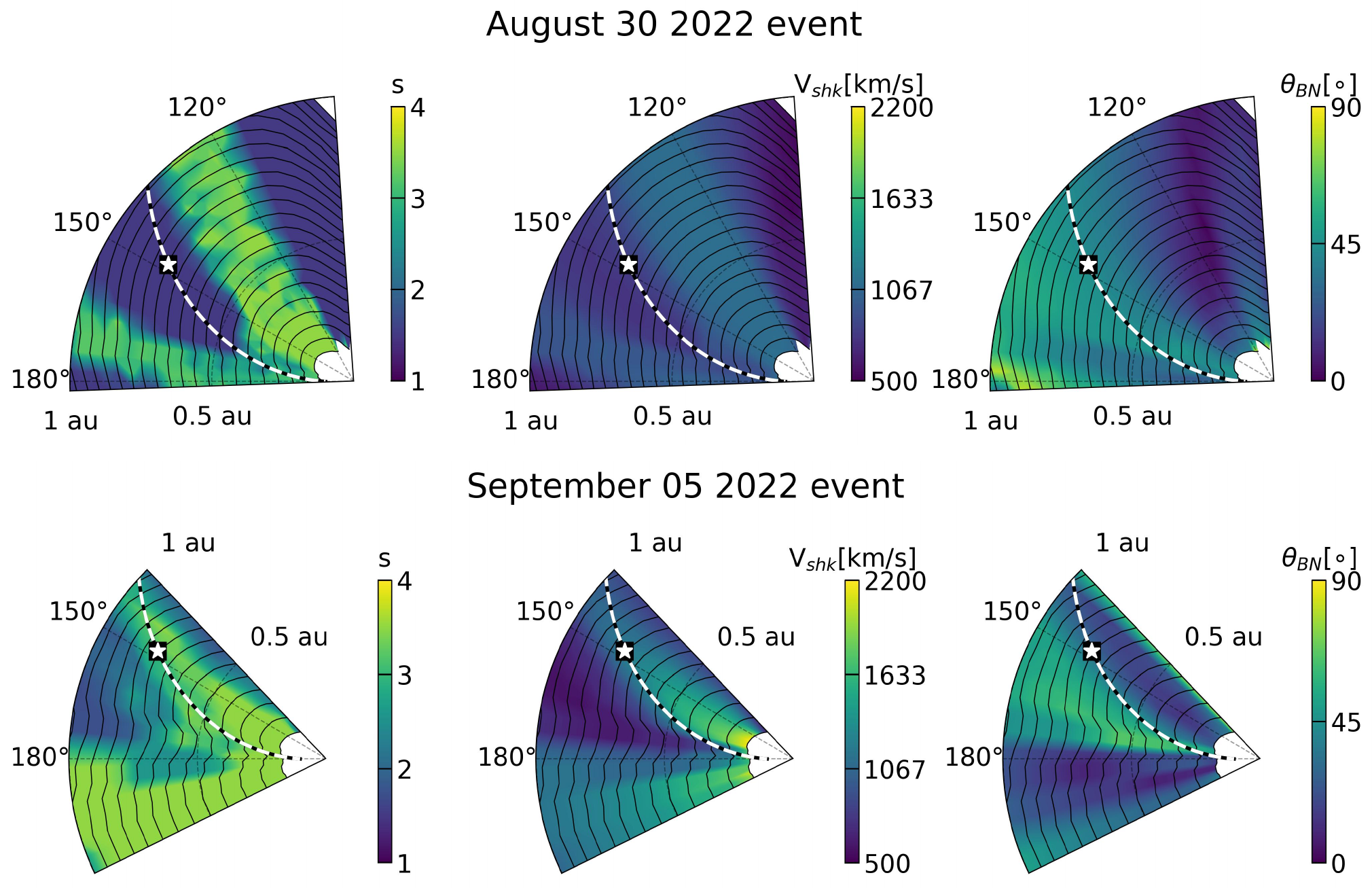}
\caption{The evolution of shock location and shock parameters (compression ratio, shock speed and obliquity angle) in the equatorial plane from $0.1$ au to $1.0$ au. The black solid curves show the shock front at different time steps. The color schemes are for different shock parameters along the shock front. The white dashed curves signal the Parker magnetic field lines passing through SolO. }
\label{fig9:shock paramters}
\end{figure*}

To illustrate the history of the shock profile, Figure~\ref{fig9:shock paramters} plots  the time evolution of shock parameters in the equatorial plane. The three panels from left to right are the shock compression ratio, the shock speed, and the shock obliquity. In each panel, the black curves represent shock fronts at various times from $0.1$ au to $1.0$ au, with the color scheme indicating the magnitude of the corresponding shock parameters along the shock front. It is assumed that SolO is located  in the solar equatorial plane as its latitude is $-3^{\circ}$. Additionally, the white dashed curves represent the Parker field lines passing through SolO at the onset of the event, i.e., the left-most vertical dashed lines in Fig.~\ref{fig1:time intensity profile}. Consider the left panels for the compression ratio. In the August event, SolO was consistently connected to shock regions  having a low compression ratio ($s<3$), even though it is a head-on event. Conversely, in the September event, SolO connects to regions with high compression ratios ($s>3$). This difference in the compression ratio is in line with the spectral index observed. The August event exhibits a softer spectral index compared to the September event. The middle panel depicts the history of shock speed. The September event shows a speed greater than $2000$ km/s near the inner boundary, whereas the August event has a lower speed of about $1000$ km/s. As discussed earlier, an important feature is the significant deceleration in the September event, which leads to the larger radial extension of the shell. Finally, the right panels plot the shock obliquity angle $\theta_{\rm BN}$. In the August event, we observe a smooth transition of shock geometry from quasi-parallel to quasi-perpendicular as the connection moves from the eastern flank to the west flank.  The September event shows a large shock obliquity angle in the nose of the shock, resulting from the depression caused by a slow stream ahead of the shock.

\subsection{Time profiles and spectra from iPATH model for both events}\label{subsec:SEP_ipath }

\begin{figure*}
\includegraphics[width=18.9cm]{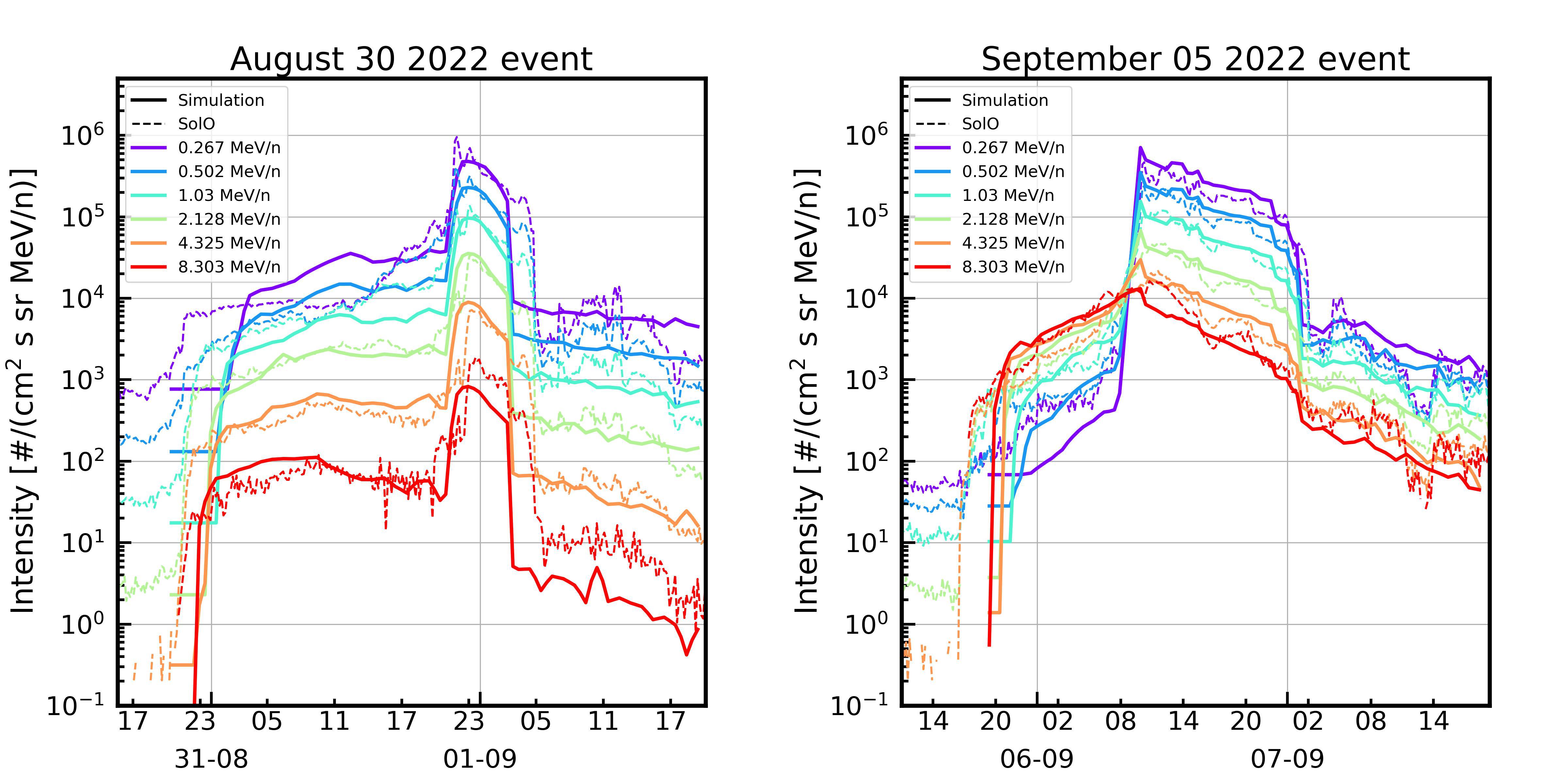}  
\caption{Time intensity profiles from the SolO observation (dashed lines) and the model calculation (solid lines). The left panel represents  the August event and  the right panel represents the September events. }
\label{fig10:modelled time profiles}
\end{figure*}

\begin{figure*}
\includegraphics[width=18.9cm]{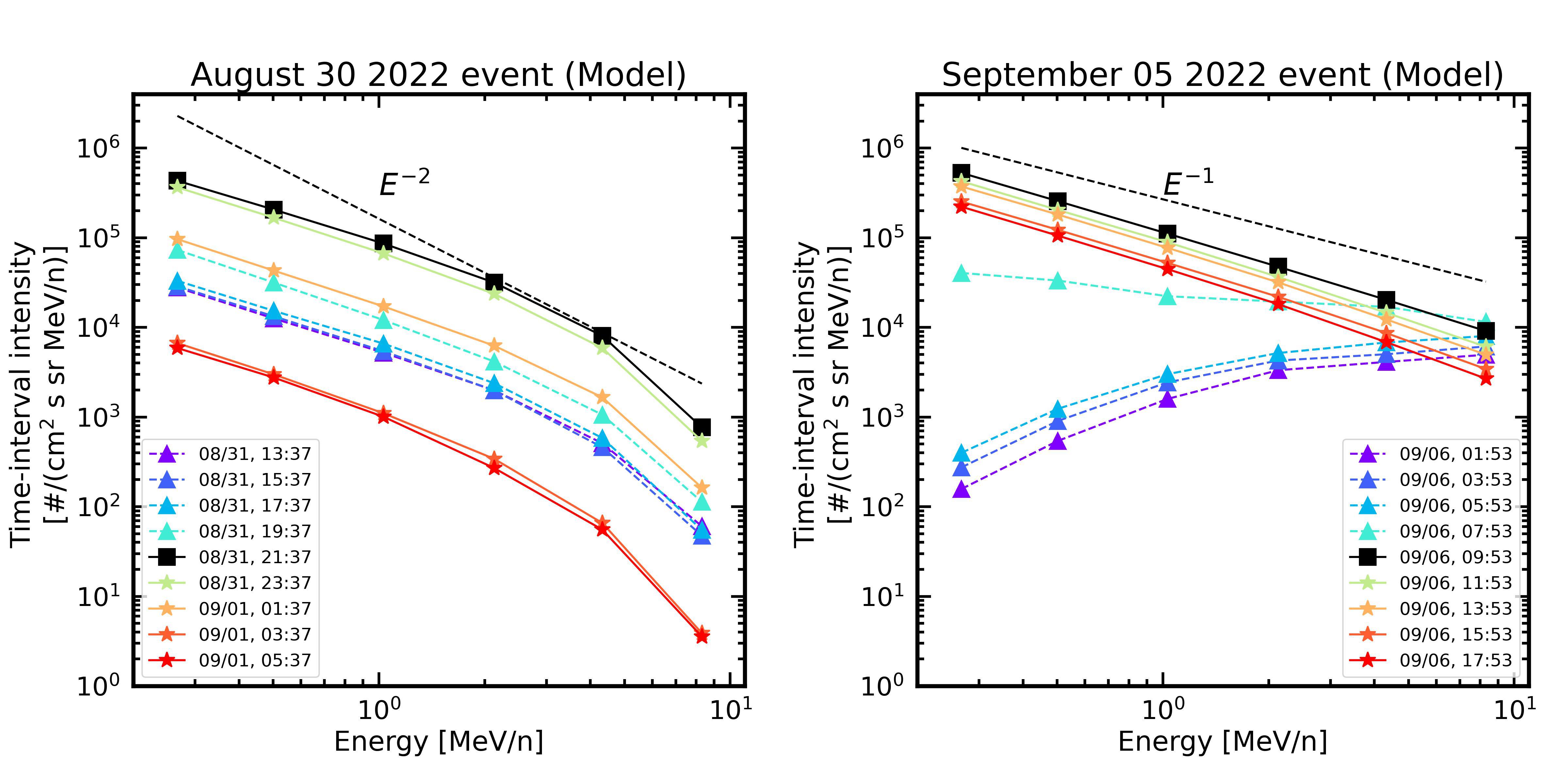}
\caption{2-hour time-interval proton spectra from the model calculation, same formats as Fig.~\ref{fig2:SIS spectra}. The left panel represents the August event and  the right panel represents the September events. }
\label{fig11:modelled spectra}
\end{figure*}

After calculating the accelerated particle spectra at the shock using equation~(\ref{eq:fp}), we use equation~(\ref{eq:1d_fp_solution_L1_Ldiff}) to obtain the escaped particle spectra at a series of times.
We track the transport of these particles and obtain the corresponding time intensity profiles and spectra at the location of SolO. Figure~\ref{fig10:modelled time profiles} shows both the observed (dashed lines) and modelled (solid lines) 48-hr proton time profiles at SolO after the CME eruption. Two critical aspects of this simulation are to account for the observed ``cross-over'' in the time profiles for the September event, as well as to reproduce the duration of the ESP phases for both events. 

With the shell module in the iPATH model, we successfully reproduced the durations of the ESP phases for both events.
The simulated duration of the ESP phase for the August and September events are approximately $5$ hours and $15$ hours, respectively. This is to be compared with observations of $7$ hours and $16$ hours. The agreement is remarkable, and to our knowledge, this is the first attempt of comparing the duration of the ESP phase between simulations and observations although \cite{Li+2003} has already examined the shell structures using a 1D iPATH simulation. 
These results highlight the necessity of using realistic MHD simulations and the shell model in better understanding ESP events. Additionally, we also model the decay of proton intensity in the sheath region, attributed to diffusion and convection resulting from the expansion of the shells, as described by \citet{Zank+etal+2000}.

As shown in Figure~\ref{fig6:fitted kappa}, the parameter $\alpha$ of escape length scale ($L_{\rm esc}= \alpha \kappa_{xx}/ U_1$) for these two events has very different energy dependence. 
We examine how this energy dependence may affect 
the observed time profile. Using Figure~\ref{fig6:fitted kappa} and observations at 0.7 au as a guide, we assume
\begin{equation}
    \alpha^{\rm Aug}(E) \sim \alpha_{0}^{\rm Aug}(E/E_0)^0. \quad \alpha^{\rm Sept}(E) \sim \alpha_{0}^{\rm Sept}(E/E_{0})^{-0.4}.
    \label{eq:alphaChoice}
\end{equation}
For both events, the  reference energy $E_0 = 1.0$ MeV/n. We choose $\alpha_{0}^{\rm Aug}=3$  for the August event and $\alpha_{0}^{\rm Sept}=5$. 
Note that the energy dependence of $\alpha$ may change as a function of radial distance. However, since we  only have in-situ observations at $r=0.7$ au, we do not explore the radial dependence of
$\alpha$ in this work.

Our choices of $\alpha$ in equation~(\ref{eq:alphaChoice}) effectively simulate the observed behavior of particle intensity in both events. For the August event, fitting in Fig.~\ref{fig6:fitted kappa} suggests a weak energy dependent of $L_{\rm esc}$, leading to our  choice of an energy-independent $\alpha$ in equation~(\ref{eq:alphaChoice}).  Particles escape from $L_{\rm esc}=3L_{\rm diff}$ upstream of the shock. This choice is similar to previous modeling efforts using iPATH \citep{Zank+etal+2000,Li+2003,Hu+etal+2017}. The results, as presented in Fig.~\ref{fig10:modelled time profiles}, demonstrate that the modeled time profiles reasonably match the observed data. The intensity exhibits a similar magnitude of decay upstream of the shock, a consequence of the assumption of energy-independent $\alpha$.
For the September event,  the choice of $\alpha$ in equation~(\ref{eq:alphaChoice}) leads to the appearance of the cross-over of the time profiles, which qualitatively match the observations. This suggests that the cross-over is a consequence of a large $\alpha_0$ and a stronger energy dependence $(E/E_0)^{-0.4}$. This is one important result of the present work. 
Note that for the lowest energy channel of $0.267$ MeV/n, the modelled intensity is notably lower than the measurements for the September 5 event, suggesting that these particles are trapped and barely escape from the shock in our simulation. This implies that we overestimate the escape length for $0.267$ MeV/n protons, perhaps if the energy dependence of $\alpha$ is not a single power law as we assumed. It is worth noting that whether cross-over can be observed in time profiles is also related to the spectral index of the accelerated particles. It is easier for a cross-over to occur for a harder spectrum than for a softer spectrum. This is because the intensity difference between two energies, e.g. $E_1$ and $E_2$, is larger (smaller) for a softer (harder) spectrum.
Therefore, the occurrence of ``cross-over'' is more likely to occur in events with strong shocks. 

We emphasize that the selection of the escape length scale $L_{\rm esc}$ for these two events is based on the in-situ observations at $r \sim 0.7$ au and we assume it has no radial dependence. This suggests that the exact value of $L_{\rm esc}$ is not known a priori and may vary for different events. Further investigations that aim to better understand how $L_{\rm esc}$ vary in different events should be pursued.

Fig.\ref{fig11:modelled spectra} plots the simulated particle spectra at intervals of 2 hours, from 8 hours before the shock passage to 8 hours after the shock passage. The labels indicate the corresponding start times for each spectrum. To ensure a  direct comparison with the observed spectra in Fig.\ref{fig2:SIS spectra}, the simulated spectra in Fig.\ref{fig11:modelled spectra} are presented in the same format. In the case of the August event, the modelled spectra exhibit a slightly harder spectrum above $1$ MeV/n compared to the observations. This discrepancy may arise from the fact that the simulated compression ratio is slightly high. For the September event, our model successfully reproduces similar positive spectral indices upstream shock and the negative indices $\sim -1$ downstream shock, which are also consistent with the observations. The proper choice of the escape length scale plays a crucial role in understanding this ESP event. If a larger escape length scale is employed, we observe steeper spectra with positive spectral indices. This event serves as an example highlighting the significance of upstream turbulence in controlling particle acceleration and escape processes.

\section{Summary and Conclusion}\label{sec:conclu}

In this study, we have investigated two ESP events observed by Solar Orbiter on August 30 2022 and September 5 2022. These two events exhibited different characteristics in terms of ESP duration, time intensity profiles, and spectral slope.  The ESP duration of the September event was observed to be significantly longer ($\sim 16$ hours) compared to the August event ($\sim 7$ hours). This discrepancy in duration is attributed to the passage of the shock-sheath structure associated with each event. We also compared the proton time intensity profiles and spectra for both events. In the September event, an interesting phenomenon was observed where, prior to the arrival of the shock, the flux of low-energy protons is lower than those of high-energy protons, leading to a ``cross-over'' in time profiles and positive spectral indices. Such a behavior is uncommon in ESP events and only one case was previously reported in  \citet{lario2021comparative}. In that event, the authors associated this feature with the presence of a pre-existing ICME structure. However, in our case, there was no significant magnetic cloud detected upstream of the shock. Instead,  long duration and intense magnetic fluctuations were observed, indicating the presence of strong turbulence ahead of the shock.  We calculated the power spectral density of the magnetic fluctuations with a 2-hour resolution prior to the shock passage. The PSD analysis revealed very different decay rates of PSD and spectral slopes for the two events. The September event had a long duration ($\sim 6$ hours) of enhanced turbulence and a steeper spectrum with a spectral index of $-2$.  We found that the longer duration of enhanced turbulence upstream of the shock is crucial for the observed ``cross-over'' feature of the  time profile in the September event.

Properly understanding the cross-over requires one to recognize that there are two length scales which control the behavior of the upstream particle distribution function. One length scale is the diffusion scale $L_{\rm diff}=\kappa(x_0,p)/U_1$, with $\kappa(x_0,p)$ the diffusion coefficient at the shock. It provides an estimate of the decay length scale of particle intensity upstream of the shock. Another scale is the length scale of the upstream turbulence itself, which we denote as $L_{esc}$ in this work.  Assuming the turbulence also decays exponentially, which 
leads to an exponential increase of $\kappa$ upstream of the shock, then the upstream particle distribution function is given by equation~(\ref{eq:1d_fp_solution_exp_waveintensity}) where the competition of these two length scales is clearly seen
through the parameter $\alpha$ which is 
the ratio of the escape length scale to the 
diffusion length ($L_{\rm esc}= \alpha L_{\rm diff}$). The values of $\alpha$ for these two events differ significantly. Furthermore they also have very different energy dependence. This difference in $\alpha$ leads to the  phenomenon of cross-over and an upstream particle spectrum with a positive spectral index.  

We also utilize the combined EUHFORIA and iPATH models to simulate these two events. EUHFORIA provides a good fit of the solar wind density and speed to the in-situ measurement, indicating that it captures the dynamic variation of CME deceleration. When EUHFORIA output is fed into the shell module of the iPATH code, we are able to obtain reasonable fits to the duration of ESP events. This is a  consequence of  the underlying relationship between the radial width of the individual shell and the deceleration of the CME. In the September event, a strong deceleration of CME leads to a larger radial width of the shell, thus the duration of ESP  events.   Furthermore, by choosing a large energy-dependent escape length scale based on the SolO observation, we successfully reproduced the cross-over time profiles  and  the positive spectral indices observed in the September event. These findings highlight the importance of the interplay between two length scales $L_{\rm diff}$ and $L_{\rm esc}$: one characterizing the decay of accelerated particles upstream of shock, and the other characterizing the duration of the enhanced upstream turbulence power, associated with the excitation of Alfv\'{e}n waves by streaming protons upstream of the shock. 

Our main findings are summarised as follows:
 \begin{enumerate}
 \item  The combined EUHFORIA-iPATH model, employing a realistic description of the solar wind,  provides a reasonable estimate of the duration of the ESP phase for both events.  This duration is a consequence of the deceleration history of the CME and the shock it drives,  and requires no free parameters in the model. 
\item The observed ``cross-over'' feature in the time profiles upstream of the shock in the September 05 event highlights the importance of the duration of  enhanced upstream turbulence in regulating the behavior of particle distribution function upstream of the shock.  A longer duration  of enhanced turbulence upstream of the shock, as in the September event, effectively  hinders the escape of lower-energy particles from the shock, leading to the ``cross-over''  phenomena. A criterion for having ``cross-over''  is encapsulated in the parameter $\alpha$ defined in equation~(\ref{eq:alpha_normalization}). 
``cross-over'' inevitably leads to positive spectral indices of the particle distribution functions upstream of the shock. 
\end{enumerate}

It is worth pointing out that other mechanisms beyond DSA may also be responsible for accelerating particles in SEP events.  In this study, we only consider the role of DSA, however, several recent studies have shown that particles can be accelerated up to several MeV/n via stochastic magnetic reconnection in dynamical small-scale magnetic islands downstream of shocks and/or inside magnetic cavities \citep{Zank2015ApJ...814..137Z,Khabarova2015ApJ...808..181K,Khabarova2016ApJ...827..122K,Khabarova2017ApJ...843....4K,Malandraki2019ApJ...881..116M}. Particles energized via DSA can be trapped by the magnetic islands in the shock sheath. As a result, these energetic particles can be re-accelerated to higher energies and consequently contribute to variations of time intensity profiles during the ESP phase \citep{Khabarova2021SSRv..217...38K}. A detailed examination of particle acceleration at ESP events, and in particular downstream of the shock, by including such processes should be pursued in future work.

In summary, our findings emphasize that the behavior of upstream turbulence can largely affect particle escape in SEP events. To gain a comprehensive understanding of large SEP events, SEP simulations considering realistic solar wind, CME, and ambient turbulence are crucial. 
A future statistical study examining the effect of the duration of enhanced turbulence upstream CME-driven shocks, through the parameter $\alpha$, will be pursued.

\begin{acknowledgements}
SolO in-situ data are publicly available at NASA’s Coordinated Data Analysis Web (CDAWeb) database (\url{https://cdaweb.gsfc.nasa.gov/index.html/}). 
This work is supported in part by NASA grants 80NSSC19K0075, 80NSSC22K0268,
80NSSC21K1327, and NSF ANSWERS 2149771 at UAH-USA. GL also acknowledges support through ISSI team 469. Solar Orbiter is a mission of international cooperation between ESA and NASA, operated by ESA. The Suprathermal Ion Spectrograph (SIS) is a European facility instrument funded by ESA under contract number SOL.ASTR.CON.00004.  We thank ESA and NASA for their support of the Solar Orbiter instruments whose data were used in this paper.  Solar Orbiter post-launch work at JHU/APL is supported by NASA contract NNN06AA01C and at CAU by German Space Agency (DLR) grant \# 50OT2002. The UAH-Spain team acknowledges the financial support by the Spanish Ministerio de Ciencia, Innovación y Universidades FEDER/MCIU/AEI Projects ESP2017-88436-R and PID2019-104863RB-I00/AEI/10.13039/501100011033. SP acknowledges support from the projects C14/19/089  (C1 project Internal Funds KU Leuven), G.0B58.23N and G.0025.23N (WEAVE)   (FWO-Vlaanderen), 4000134474 (SIDC Data Exploitation, ESA Prodex-12), and Belspo project B2/191/P1/SWiM. For the computations we used the infrastructure of the VSC–Flemish Supercomputer Center, funded by the Hercules foundation and the Flemish Government–department EWI.

\end{acknowledgements}

%
  \bibliographystyle{aa} 
  \bibliography{solo} 
%

\end{document}